\documentclass{article}

\usepackage{arxiv}
\usepackage{amsmath}
\usepackage[utf8]{inputenc} 
\usepackage[T1]{fontenc}    
\usepackage{hyperref}       
\usepackage{url}            
\usepackage{booktabs}       
\usepackage{amsfonts}       
\usepackage{nicefrac}       
\usepackage{microtype}      
\usepackage{cleveref}       
\usepackage{lipsum}         
\usepackage{graphicx}
\usepackage{natbib}
\usepackage{doi}
\usepackage[table,xcdraw]{xcolor}

\title{Eclipse Dynamics and X-ray Burst Characteristics in the Low-Mass X-ray Binary EXO 0748-676}

\usepackage{authblk}

\newbox{\orcid}\sbox{\orcid}{\includegraphics[scale=0.06]{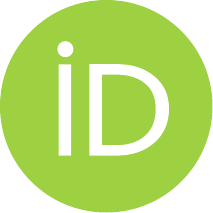}} 
\author[1]{%
	\href{https://orcid.org/0000-0001-7136-7270}{\usebox{\orcid}\hspace{1mm}Nirpat Subba\thanks{\texttt{nirpat.subbagmail.com}}}%
}
\author[2]{%
	\href{}{\hspace{1mm}Nishika Subba\thanks{\texttt{subbanishika04@gmail.com}}}%
}
\author[1]{%
	\href{}{\hspace{1mm}Jyoti Paul\thanks{\texttt{jyotipaul445@gmail.com}}}%
}
\author[3]{%
	\href{}{\hspace{1mm}Pankaj Sharma\thanks{\texttt{spankaj655@gmail.com}}}%
}
\author[4]{%
	\href{https://orcid.org/0000-0002-9566-4955}{\usebox{\orcid}\hspace{1mm}Monika Ghimiray\thanks{\texttt{ghimiraymonika@gmail.com}}}%
}

\affil[1]{Cooch Behar Panchanan Barma University, Cooch Behar, West Bengal, 736101, India}
\affil[2]{Independent Researcher, M.Sc in Physics, Sikkim Manipal University, Majitar, East Sikkim, Sikkim, 737136, India}
\affil[3]{Independent Researcher, M.Sc in Physics, North Bengal University, Siliguri, Darjeeling, WB, 734013, India}
\affil[4]{National Center for Nuclear Research, ul. Pasteura 7, 02-093 Warsaw, Poland}

\renewcommand{\shorttitle}{}

\begin{document}
\maketitle

\begin{abstract}
This study investigates the timing and spectral characteristics of X-ray bursts from the neutron star system EXO 0748-676 using the NuSTAR observatory's FPMA and FPMB instruments in the 3-79 keV range.  A notable type I X-ray thermonuclear burst at \(X = 18,479.97\), with a peak of 1,615.88 counts per second, was observed, revealing key characteristics of helium ignition on the neutron star’s surface. Distinct energy dependencies were found, with soft X-ray emissions (3-7 keV) associated with the cooling phase and mid-energy bands (7-18 keV) capturing the bulk of the energy during ignition. Hard X-rays (18-79 keV) were linked to the early stages of the explosion. The study also examined the timing of bursts, uncovering correlations between burst intervals and the neutron star’s optical period, suggesting a link between accretion processes and the system's binary nature. Analysis of the eclipse profile revealed energy-dependent ingress and egress durations, providing insights into the density gradient of the surrounding material. By applying two spectral continuum models, the research illuminated the complex emission mechanisms, with one model highlighting Comptonization and the other emphasizing both thermal and non-thermal components. Comparisons with historical data suggest temporal changes in the accretion environment, reflected in variations in photon indices and cutoff energies. The calculated flux in a full energy range of 3-79 KeV using both the continuum models are found to be  \( \sim 381.17 \times 10^{-12} \; \text{erg cm}^{-2} \text{s}^{-1} \). Calculated luminosities derived from different distance estimates range from \(  2.29 \times 10^{36}  \; \text{erg/s} \) to \(  3.86 \times 10^{36}  \; \text{erg/s} \). The findings contribute to a deeper understanding of accretion processes, thermonuclear dynamics, and the behavior of neutron stars, with implications for future investigations into the complex interactions in low-mass X-ray binaries and high-energy astrophysical phenomena.
\end{abstract}


\section{Introduction}
\label{intro}

The idea of \textbf{low-mass X-ray binaries (LMXBs)} \cite{liu2001catalogue} emerged in the late 1960s and early 1970s, soon after X-ray sources were discovered in space. Research on binary systems involving compact objects like neutron stars or black holes began with the identification of bright X-ray sources by satellites such as \textbf{Uhuru}, the first satellite dedicated to X-ray observations, launched in 1970 \cite{riccardo1962}.
The groundbreaking work that identified X-ray binaries as a unique category of astronomical objects is largely attributed to \textbf{Riccardo Giacconi}, \textbf{Herbert Gursky}, and their colleagues. Giacconi, who received the \textbf{Nobel Prize in Physics} in 2002, was instrumental in both the discovery of X-ray sources and the advancement of X-ray astronomy.Additionally, high-mass X-ray binaries (HMXBs) produce strong and variable X-ray emissions due to interactions with the winds from their massive companion stars \cite{charles2003optical}, while low-mass X-ray binaries (LMXBs)\cite{liu2001catalogue} usually exhibit more stable X-ray emissions because the material transfer from their smaller companion stars is more consistent. Studying both types of binary systems is important because findings from one can help us understand the other, providing a clearer picture of how these binary star systems evolve over time \cite{kahabkaVenDen1997luminous}.\\

EXO 0748–676 is a well-known neutron star (NS) in a low-mass X-ray binary (LMXB) system that has garnered attention for its frequent and diverse thermonuclear (type I) X-ray bursts. Discovered by the \textbf{European X-ray Observatory Satellite (EXOSAT)}, EXO 0748-676 in February 1985 \cite{parmar1986discovery}. EXO 0748-676 is recognized for producing both type I X-ray bursts, caused by thermonuclear explosions on the neutron star's surface, and continuous X-ray emissions from the accretion process \cite{gottwald1987}. These phenomena provide valuable insights into the behavior of matter under extreme conditions. several studies have focused on the binary system EXO 0748-676, utilizing NuSTAR \cite{harrison2013nuclear} and XMM-Newton \cite{bhattacharya2024xmm} to better understand its complex behaviour. Recent NuSTAR observations in 2024 captured high-energy phenomena, such as Type I X-ray bursts and consecutive eclipses, shedding light on the accretion processes around the neutron star and the system's post-quiescence evolution. In parallel, other research combined NuSTAR's data with XMM-Newton’s high-resolution spectroscopy  \cite{bhattacharya2024xmm} to analyze changes in emission lines before, during, and after bursts. These studies highlighted how circumstellar material and physical processes evolve during outbursts, deepening insights into the neutron star’s environment. Together, these findings enhance our understanding of the system’s thermonuclear bursts and eclipsing features. \\

EXO 0748-676 is notable for being the only source where gravitationally redshifted absorption lines during X-ray bursts have been observed \cite{cottam2002,luo2013redshift}. Identified redshifted spectral lines corresponding to O and Fe transitions during X-ray bursts from EXO 0748-676 ,using the Reflection Grating Spectrometer(RGS) instrument on XMM-newton, indicated a gravitational redshift of z=0.35 at the neutron star's surface, corresponding to a mass-to-radius ratio of $M/R=0.152 M_\circ/km$. This finding provides an important empirical constraint on the equation of state (EoS) for dense, cold nuclear matter \cite{cottam2002}. \\

In 1985, EXO 0748-676 was found to show dips and eclipses every 3.82 hours \cite{parmar1986discovery,southwell1996optical,degenaar2014probing}, with each eclipse lasting 8.3 minutes \cite{psaradaki2018modelling}. These periodic X-ray dips and eclipses were detected by EXOSAT, indicating that material is regularly blocking the X-ray source, likely due to interactions between the companion star and the accretion disk around the neutron star. These features suggest that EXO 0748-676 is viewed at an angle of about $75^{\circ}-83^{\circ}$ \cite{parmar1986discovery}, almost aligned with the plane of the accretion disk. There are several low-mass X-ray binaries (LMXBs)\cite{liu2001catalogue} with high inclinations, but only a few of them exhibit eclipses. Notable examples include EXO 0748-676 \cite{parmar1986discovery} and MXB 1659-298\cite{sidoli2001xmm}.\\

Due to the high inclination of EXO 0748-676, the observer’s viewpoint is nearly aligned with the accretion disk, making it ideal for analyzing structures above the disk via spectral lines. High-resolution spectra from Chandra/HETG \cite{jimenez2003discrete} and XMM-Newton/RGS have been previously used to study this system. Cottam \cite{cottam2001high}compared spectra during different phases, such as low-emission periods, rapid variations, and bursts, detecting broadened recombination lines from nitrogen, oxygen, and neon. Additionally, T.M.S. Kallman and his team found a gravitationally redshifted absorption feature (z = 0.35) in the X-ray spectrum of EXO 0748-676 \cite{cottam2002}, which is important for determining key neutron star properties, like the ratio of its radius-to-mass \cite{bhattacharyya2006shapes} and exploring the superdense matter within the neutron star core \cite{bhattacharyya2010measurement}.\\

This study thoroughly investigates EXO 0748-676, a neutron star located in a Type I X-ray binary system, intending to uncover its distinct features through detailed timing and spectral analyses. Data from NuSTAR observations were meticulously analyzed to gain insights into the unique behavior and properties of the source. The paper is structured as follows: the introduction in section \ref{intro} provides an overview of the study and its significance; section \ref{obs} explains the data reduction process from NuSTAR; section \ref{result} presents the results, which are further divided into two subsections: timing analysis in subsection \ref{time} and spectral analysis in subsection \ref{spectral}. The conclusion is summarized in section \ref{conc}, followed by data availability information in section \ref{availability} and a list of references. This structure allows readers to navigate the study easily and understand the various components of the investigation.



\section{Observation and data reduction}
\label{obs}

NuSTAR (Nuclear Spectroscopic Telescope Array), launched by NASA in 2012, \cite{harrison2013nuclear} is the first telescope capable of focusing high-energy X-rays (3-79 keV). It provides unprecedented clarity for studying energetic events such as black holes, neutron stars, and supernova remnants. NuSTAR's advanced technology uses X-ray mirrors and detectors to capture detailed images of high-energy environments, complementing other observatories like Chandra \cite{2009chandra} and XMM-Newton \cite{bhattacharya2024xmm}. Its significant contributions include detecting X-ray jets from black holes, studying supernova remnants, and deepening our understanding of the physics around compact objects like neutron stars. NuSTAR (Nuclear Spectroscopic Telescope Array) recently observed EXO 0748-676 on June 17, 2024, recording an average net count rate of 5.0 counts per second (in the 3-79 keV energy range, with a 2 arcminute extraction radius and both focal plane modules combined). \\

\begin{figure*}
    \centering
    \rotatebox{270}{\includegraphics[width=.7\linewidth]{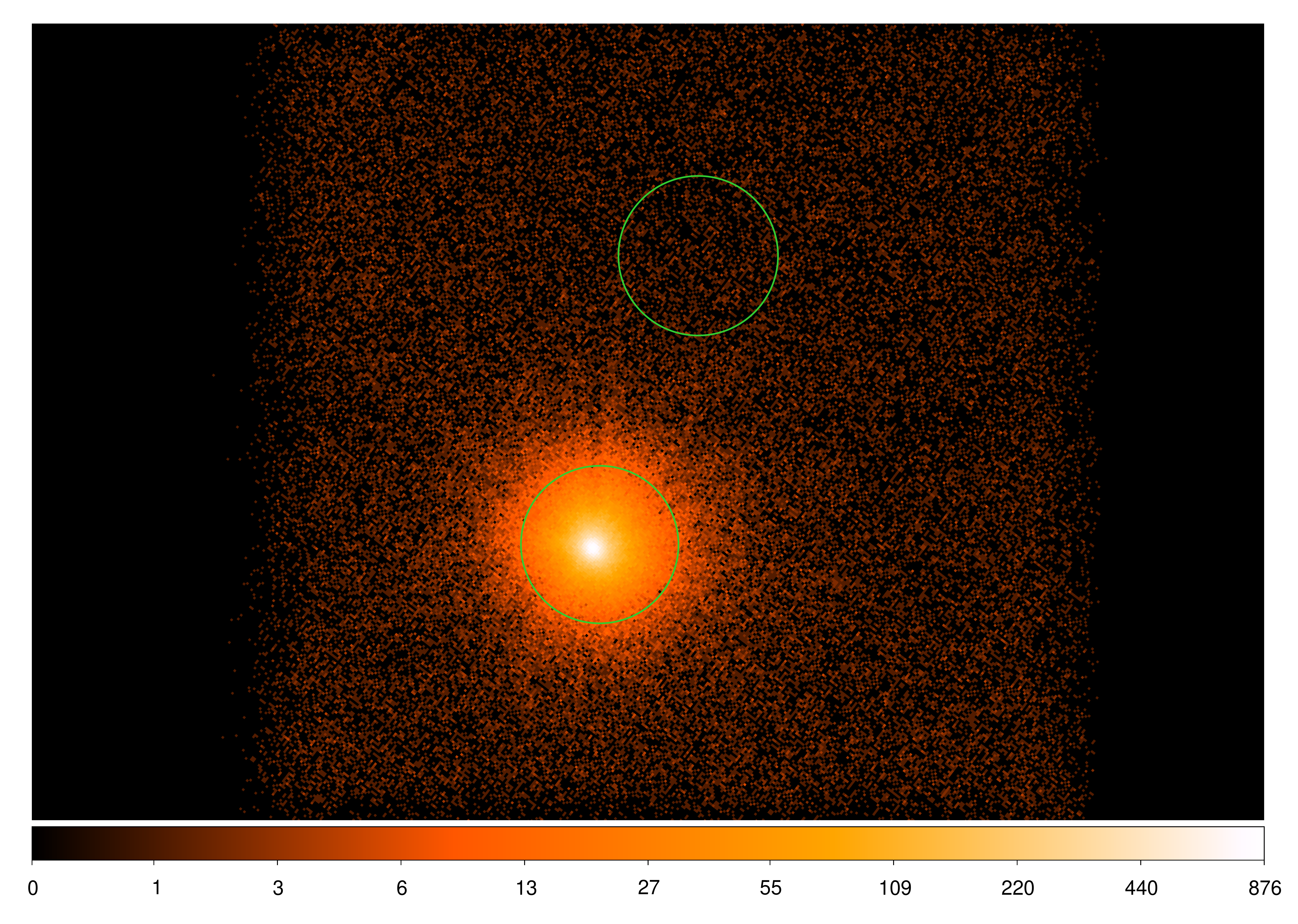}}
    \caption{The figure displays the selected regions for the source and background in the NuSTAR observation, associated with Observation ID 91001324002.
}
    \label{fig:1}
\end{figure*}

The lightcurve for the source EXO 0748-676 is shown in Fig. \ref{fig2:light-curve-brust}, while the detailed information regarding the observation IDs and their respective exposure times is provided in Table \ref{tab:1}.
\newline The data processing for EXO 0748-676 was carried out using HEA-SOFT 6.34 includes XSPEC v12.14.1 and CALDB v20240930 for NuSTAR. Light curves, spectra, response matrices, and effective area files were generated with NUSTARDAS software v2.1.4. Unfiltered event files were cleaned using the mission-specific tool NUPIPELINE. A circular source region of 100" was defined around the source center, with a background region selected away from the source (as shown in Fig. \ref{fig:1}). Light curves and spectra were extracted from the defined source and background regions using XSELECT and NUPRODUCTS commands. Background subtraction was handled separately for FPMA and FPMB data, and their light curves were combined using the LCMATH tool. The NuSTAR observations of the system were then analyzed for timing and spectral data.

 \begin{table}
     \centering
     \caption{NuSTAR observation details of EXO 0748-676}
     \vspace{0.25cm}
     \begin{tabular}{|c|c|c|}
        \hline
        \textbf{Observation} & \textbf{Observation Start Date} & \textbf{Exposure Time} \\
        \textbf{ID} & \textbf{Time(UT)} & \textbf{(ks)}\\
        \hline
        91001324002 & 2024-06-17 01:31:09 & 62.648 \\ 
        \hline
     \end{tabular}
     \label{tab:1}
 \end{table}

\section{Data analysis and Results}
\label{result}

\subsection{Timing analysis}
\label{time}

\subsubsection*{Type I X-ray burst}


In studying the NuSTAR observatory data of the low-mass X-ray binary (LMXB) \cite{liu2001catalogue} EXO 0748-676, after obtaining the time curve lc (light curve) file and the spectral data, PHA (Pulse Height Amplitude) file using the \texttt{nuproducts} command, the background noise was subtracted from the source using the \texttt{lcmath} command. This background subtraction was carried out separately for both the FPMA and FPMB observation data. Once the background was successfully removed from each observation, the two source data sets (FPMA and FPMB) were combined into a single file using \texttt{lcmath}. Afterward, barycentric correction was applied using the \texttt{barycorr} command along with the appropriate auxiliary file containing the necessary information for the correction. This process corrected the timing data to account for the Earth's motion, adjusting the light curve to the Solar System barycenter for accurate pulsar timing analysis.\\

\begin{figure*}[h!]
    \centering
    \includegraphics[width=01\linewidth]{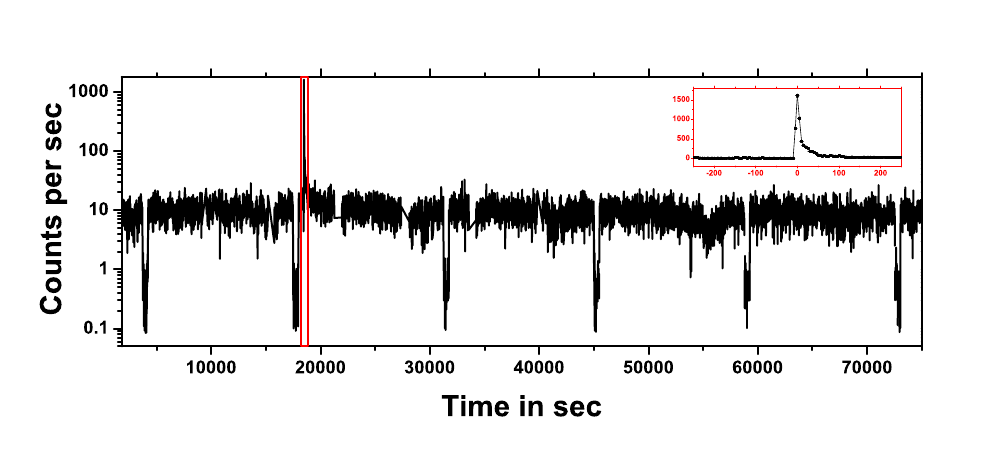}
    \caption{The figure presents the lightcurve of EXO 0748-676, a neutron star, observed during NuSTAR Observation ID 91001324002}
    \label{fig1:light-curve-brust}
\end{figure*}

After completing the previous steps, the light curve was plotted by selecting the relevant time intervals and applying binning to improve time resolution, using the \texttt{lcurve} command. Figure \ref{fig1:light-curve-brust} shows the light curve of the LMXB pulsar EXO 0748-676 in the energy range of 3–79 keV. The plot uses a logarithmic scale for the count rate axis, allowing clear visualization of the eclipse and the prominent burst events in the source EXO 0748-676. In Fig. \ref{fig1:light-curve-brust}, the red lines represent the most significant burst observed at \(X = 18,479.97\) (MJD 20478.519486587) in the neutron star EXO 0748-676 during NuSTAR observation ID 91001324002. The same burst is also shown magnified in the top-right panel to highlight its detailed structure. Clearly, from Fig. \ref{fig1:light-curve-brust}, five distinct eclipse cycles can be observed, the properties of which will be discussed in more detail in a later section.\\

\begin{figure*}[h!]
    \centering
    \includegraphics[width=01\linewidth]{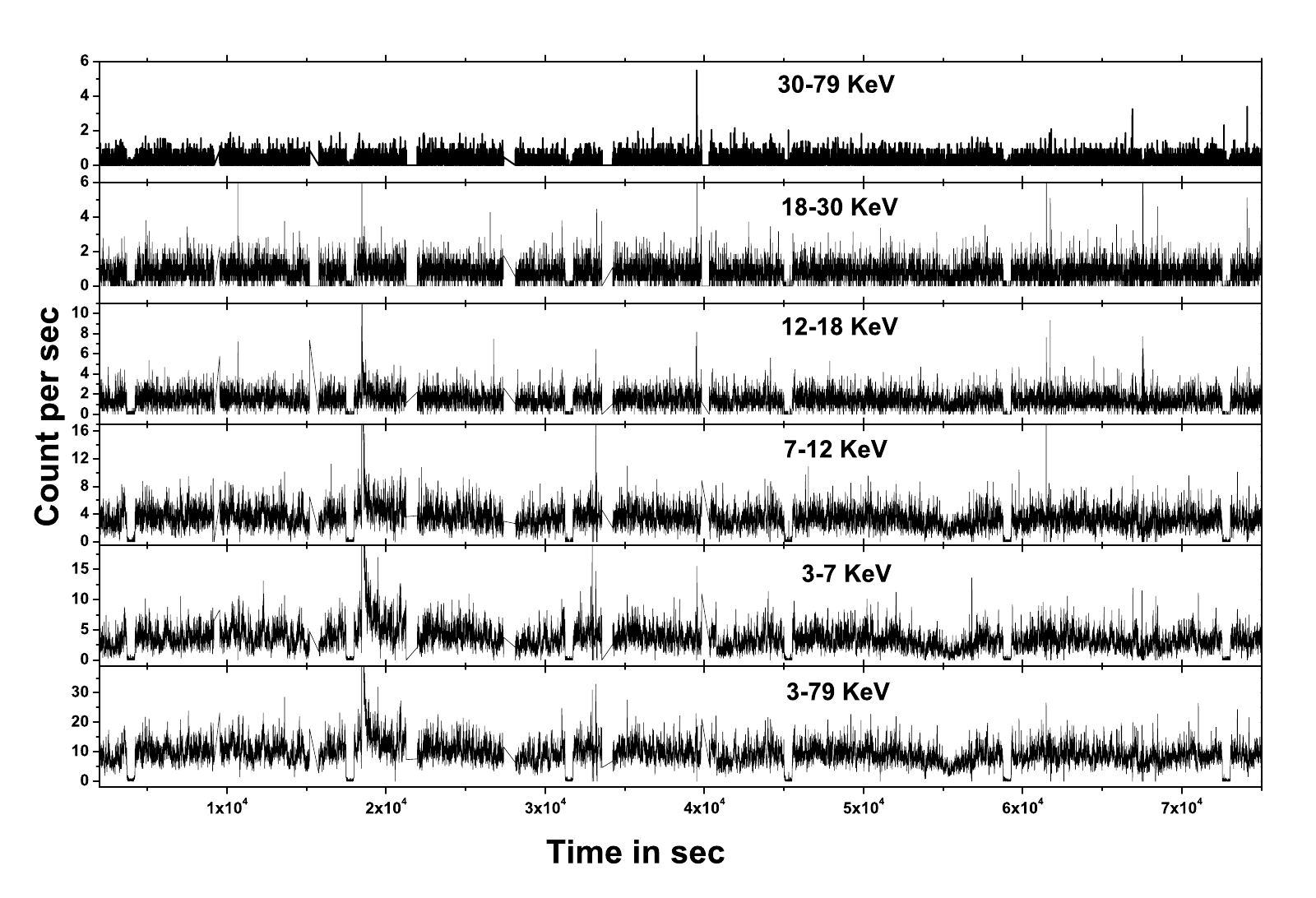}
    \caption{The figure presents the lightcurve of EXO 0748-676, a neutron star, observed during NuSTAR Observation ID 91001324002 at different energy band.}
    \label{fig2:light-curve-brust}
\end{figure*}

The light curve was then further divided into different energy bands to analyze energy-dependent variations in the count rate. Energy filters corresponding to specific ranges (e.g., 3–7 keV, 7–12 keV, etc.) were applied to split the light curve into segments, allowing for a detailed study of the source's behavior across different energy ranges and detecting burst events. This was done using the \texttt{nuproducts} command, where specific energy ranges were defined in terms of PI channels, followed by the same steps as above using \texttt{lcmath} for background subtraction and \texttt{barycorr} for barycentric correction for all the energy bins. In Fig. \ref{fig2:light-curve-brust}, the time curve for the full energy range (3–79 keV) is shown alongside smaller energy ranges (3–7 keV, 7–12 keV, and up to 30–79 keV) to examine the nature of the outburst in EXO 0748-676. This enables a detailed analysis of how the outburst evolves across different energy bands. Note that the actual height of the significant burst is clipped to make smaller outbursts more visible.\\

The analysis of X-ray bursts in the 3-79 keV energy range from the neutron star system reveals significant characteristics of Type I X-ray bursts \cite{gottwald1987}, commonly observed in low-mass X-ray binaries (LMXBs). These bursts are thermonuclear explosions resulting from the unstable burning of accreted material, predominantly hydrogen and helium, on the surface of the neutron star. Such bursts exhibit rapid rise times (0.5-5 seconds) followed by slower decays (10-100 seconds) \cite{strohmayer2003new}, releasing immense amounts of energy—on the order of \(10^{39}\) ergs in just a few seconds. This energy output is vastly greater than the Sun's weekly energy release, driven by runaway nuclear reactions ignited when sufficient material accumulates on the neutron star's surface \cite{bhattacharyya2009x}.\\

The first notable burst in this dataset occurs at \(X = 13,594.02\), with a count rate of 28.5, indicating a modest energy release. This is likely a result of localized instabilities in the neutron star’s magnetosphere or fluctuations in its accretion flow, producing only a minor deviation in the X-ray count rate. Such smaller bursts are typical of low-level accretion-driven instabilities but do not represent major thermonuclear events.\\

A dramatic shift in behavior is observed at \(X = 18,479.97\), where a major burst is detected with a count rate of 1,615.88. This burst is indicative of a powerful Type I X-ray burst, caused by thermonuclear fusion on the neutron star’s surface \cite{galloway2008biases}. Type I bursts occur when accreted material from a companion star accumulates on the neutron star’s surface, reaching critical density and temperature. At this point, a runaway nuclear reaction is triggered, releasing vast amounts of energy in a short span. This particular burst likely involved helium ignition, where the accreted matter had built up over hours or days, reaching a column density of approximately \(10^8 \, \text{g} \, \text{cm}^{-2}\) and igniting through the unstable triple-alpha process \cite{bhattacharyya2009x}. Such bursts are frequently observed in LMXBs and are marked by sudden, intense X-ray emissions as the surface material ignites in a thermonuclear explosion \cite{liu2001catalogue}.\\


This analysis of the X-ray burst data in the 3-79 energy range, combined with the neutron star's optical period of 13,767.5 seconds , reveals key aspects of the system’s behavior. It supports the interpretation that the \(X = 18,479.97\) burst represents a significant thermonuclear event, while the smaller bursts, such as the one at \(X = 13,594.02\), likely result from less energetic fluctuations related to the star’s accretion system or magnetospheric activity. These findings align with the well-documented behaviors of Type I X-ray bursts, where accretion, nuclear burning, and rotational dynamics all interplay to generate recurrent, explosive energy releases.\\

Following the main burst at \(X = 18,479.97\), another burst is detected at \(X = 19,463.97\) with a count rate of 32.14. This event likely represents a "recoil" burst, where the system releases residual energy after the primary thermonuclear explosion. It is common in neutron star systems for smaller, secondary bursts to follow the main burst as the system stabilizes. These recoil bursts are driven by remaining instabilities in the accreted material or disturbances in the neutron star's magnetic field, allowing the system to return to equilibrium.\\

Subsequent smaller bursts occur at \(X = 33,242.53\) and \(X = 35,140.60\), with count rates of 32.93 and 27.55, respectively. These bursts, although significant, are much less energetic than the primary event. They likely result from local fluctuations in the neutron star's accretion flow or minor magnetic reconnections. These types of events are characteristic of accreting neutron stars, where variations in material flow and magnetic activity can produce periodic energy releases in the form of X-ray emissions.\\

Overall, these secondary bursts, including the recoil event, reflect the system's post-burst behavior, where residual energy dissipation and localized adjustments in the accretion process or magnetosphere lead to smaller bursts of energy. While not as dramatic as the main thermonuclear explosion, these events are still important in understanding the complex dynamics of neutron star systems.\\

The intervals between these bursts provide further insights into the system’s dynamics. The gap between the first major burst at \(X = 18,479.97\) and the recoil burst at \(X = 19,463.97\) is approximately 984 seconds, reflecting rapid energy release and adjustment. More intriguingly, the gap between the burst at \(X = 19,463.97\) and the next significant burst at \(X = 33,242.53\) is roughly 13,778 seconds, closely matching the neutron star's optical period of 13,767.5 seconds. This suggests a potential link to the star's rotational cycle, where some bursts could be influenced by periodic accretion events associated with its binary interaction, pointing to the role of rotational dynamics and the magnetic field structure in modulating the burst mechanism.\\

Supporting the thermonuclear origin of these bursts, studies by Strohmayer \textit{et. al.} \cite{strohmayer2006neutron,strohmayer2003new} demonstrate that the burst emission area matches the expected neutron star surface area, confirming that the bursts stem from nuclear burning rather than accretion disk processes. Earlier theoretical predictions by Hansen and Van Horn (1975) \cite{hansen1975steady} suggested that neutron stars with thin hydrogen and helium layers would undergo such unstable nuclear burning, as stable burning would result in energy dissipation through gravitational processes.\\

These bursts follow distinct burning regimes based on the accretion rate. At moderate accretion rates, hydrogen burns steadily via the hot CNO cycle \cite{strohmayer2003new,fowler1965nucleosynthesis,bildsten2000theory}, allowing helium to accumulate and ignite in a sudden burst, referred to as a helium burst. In cases where the burst surpasses the Eddington luminosity, a photospheric radius expansion (PRE) burst may occur, where radiation pressure lifts the neutron star’s outer layers. In addition to the well-documented Type I X-ray bursts, another class of thermonuclear bursts known as superbursts has been observed. These bursts are characterized by significantly longer recurrence times, on the order of years, and differ from Type I bursts in both decay time and energy output. While Type I bursts last for seconds to minutes, superbursts exhibit much slower decay times, ranging from 1 to 3 hours, and release vastly more energy, around \(10^{42}\) ergs. The first superburst was discovered by Cornelisse et al. \cite{cornelisse2000longest}, marking a major distinction from the shorter and less energetic Type I bursts typically seen in neutron star systems.\\

\begin{figure*}[h!]
    \centering
    \includegraphics[width=0.8\linewidth]{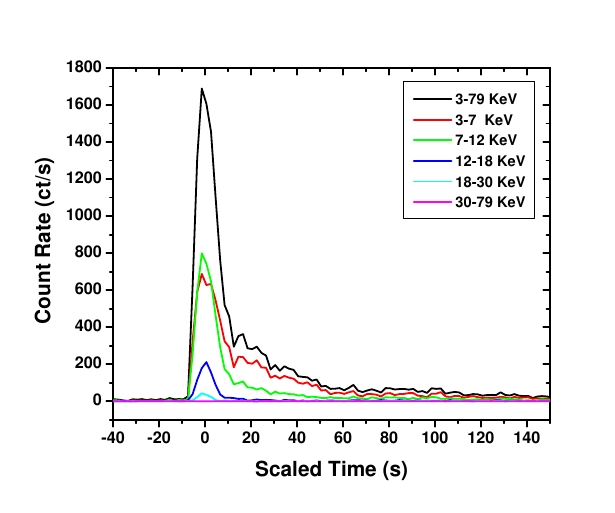}
    \caption{Light curve of EXO 0748—676 at different energy range showing a type I X-ray burst observed at MJD 20478.519486587 during NuSTAR observation ID 91001324002.  }
    \label{fig:burst_all_time}
\end{figure*}

Fig. \ref{fig:burst_all_time} shows the significant Type I X-ray burst observed at MJD 20478.519486587 during the NuSTAR observation, displayed across different energy ranges.  The energy range over which X-ray bursts are observed plays a significant role in understanding the physical mechanisms driving these thermonuclear explosions on neutron stars. In the broad 3-79 keV energy range, the burst at \(X = 18,479.97\) stands out as the most energetic event, with a count rate of 1615.88 counts per second. This range captures the full evolution of the burst, from the initial thermonuclear ignition to the cooling phase, providing a complete picture of the burst dynamics. The broad energy spectrum shows not only the intense peak but also the subsequent smaller bursts, such as the recoil burst at \(X = 19,463.97\), with a count rate of 32.14. These secondary bursts likely result from residual energy dissipation or magnetospheric adjustments. However, when we break down the burst behavior into narrower energy ranges, important patterns emerge. From Fig. \ref{fig:burst_all_time}, in the soft X-ray band (3-7 keV), the primary burst shows a much lower count rate of 687.71 counts per second, indicating that much of the burst’s energy is emitted in higher energy bands. Soft X-rays primarily capture the cooling phase of the burst, which decays more slowly, highlighting the prolonged emission from the neutron star’s surface. This behavior aligns with our understanding of Type I X-ray bursts, where initial high-energy photons rapidly decline, leaving softer X-rays to dominate the emission for longer periods \cite{galloway2008thermonuclear}.\\

In contrast, mid-energy ranges (7-12 keV and 12-18 keV) show bursts with intermediate intensity, emphasizing the importance of these bands in capturing the peak of the thermonuclear explosion. For instance, in the 7-12 keV range, the primary burst reaches 799.14 counts per second, while in the 12-18 keV range, it drops to 211.78 counts per second. This suggests that the bulk of the burst’s energy is emitted in these mid-energy bands, where nuclear burning on the neutron star's surface releases a significant amount of X-rays before cooling. The physics here is driven by helium burning, which ignites under conditions of high pressure and temperature on the star’s surface, leading to rapid energy release followed by radiative cooling \cite{strohmayer2003new,fowler1965nucleosynthesis,lewin1997x,lewin1993x}.\\

\begin{figure*}[h!]
    \centering
    \includegraphics[width=0.8\linewidth]{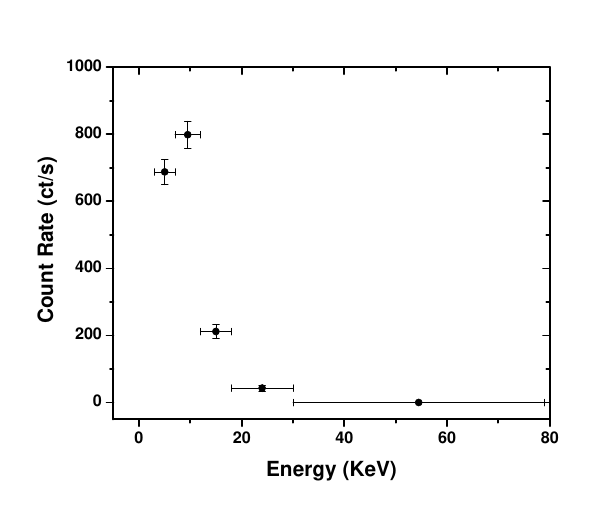}
    \caption{Variation of Type I X-ray bursts count rate with different energy range observed in LMXBs EXO 0748-676 at MJD 20478.519486587 during NuSTAR observation }
    \label{fig:burst_with_en}
\end{figure*}

Higher energy bands (18-30 keV and 30-79 keV) capture the hardest X-rays, but the intensity is much lower than the full energy range. For instance, in the 18-30 keV band, the primary burst has a count rate of 43.28318 counts per second, while in the 30-79 keV band, it further decreases to 1.87 counts per second. These hard X-rays are likely emitted during the early phases of the burst, reflecting the hottest parts of the thermonuclear explosion. The variation in the amplitude of the Type I X-ray burst count rate with energy is plotted in Fig. \ref{fig:burst_with_en}.  The rapid decline in higher energy bands supports the idea that hard X-rays are primarily linked to the initial stages of nuclear burning, with less emission as the system cools. Such behavior is consistent with the high-energy photons emitted during the peak of a photospheric radius expansion (PRE) burst, where radiation pressure lifts the outer layers of the neutron star before settling into a cooling phase \cite{strohmayer2006neutron,strohmayer2003new}.\\

\subsubsection*{Eclipse profile}

\begin{figure*}[h!]
    \centering
    \includegraphics[width=0.9\linewidth]{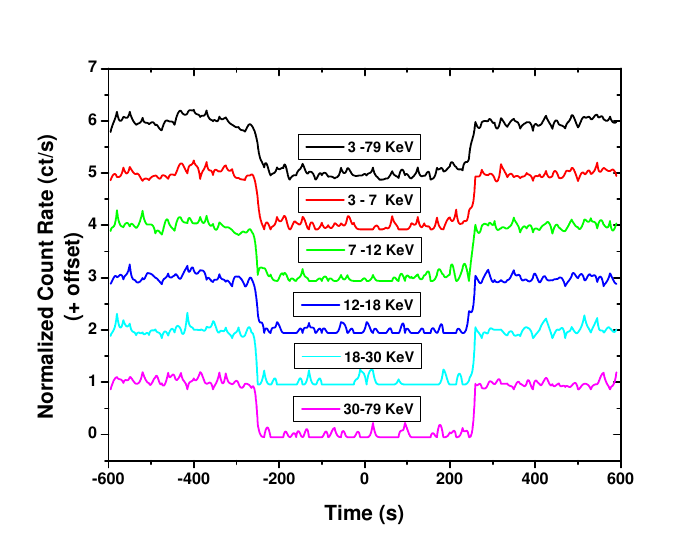}
    \caption{Normalized Eclipse profile of the LMXBs EXO 0748-676 observed at different energy ranges with verticle offset for clarity.}
    \label{fig:norm_eclipse}
\end{figure*}

From Fig. \ref{fig1:light-curve-brust}, five distinct eclipse cycles were observed in the NuSTAR observation of the LMXB EXO 0748-676. Each eclipse cycle, caused by the low-mass main-sequence secondary star UY Vol passing in front of the neutron star, has a period of \(13767 \pm 7.906\) seconds, or approximately 3.824 hours, which closely matches the previously reported orbital period of EXO 0748-676 at 3.824 hours \cite{wade1985optical,knight2022eclipse,southwell1996optical}. The eclipse typically lasts about \(496.667 \pm 2.47\) seconds, equivalent to \(8.278 \pm 0.041\) minutes. On average, the eclipse ingress/egress time is found to be about 12 seconds. To understand more about the eclipse and its energy dependency, we have folded the extracted light curve about the orbital period of 3.824 hours and normalized it throughout the light curve with the average count rate of the burst-free and out-of-eclipse light curve so that the eclipse profile now has an out-of-eclipse count rate of 1.0, and the eclipse totality leveled at 0.0. The eclipse profile is extracted for different energy ranges and is plotted in Fig. \ref{fig:norm_eclipse}. Note that the eclipse profile at different energy ranges is set to a vertical offset for visual clarity. From Fig. \ref{fig:norm_eclipse}, it is observed that initially during ingress, there is a gradual decrease in the count rate towards totality, and the same phenomenon is observed at egress, where the count rate gradually rises to the normalized count rate.

\begin{figure*}
    \centering
    \includegraphics[width=0.9\linewidth]{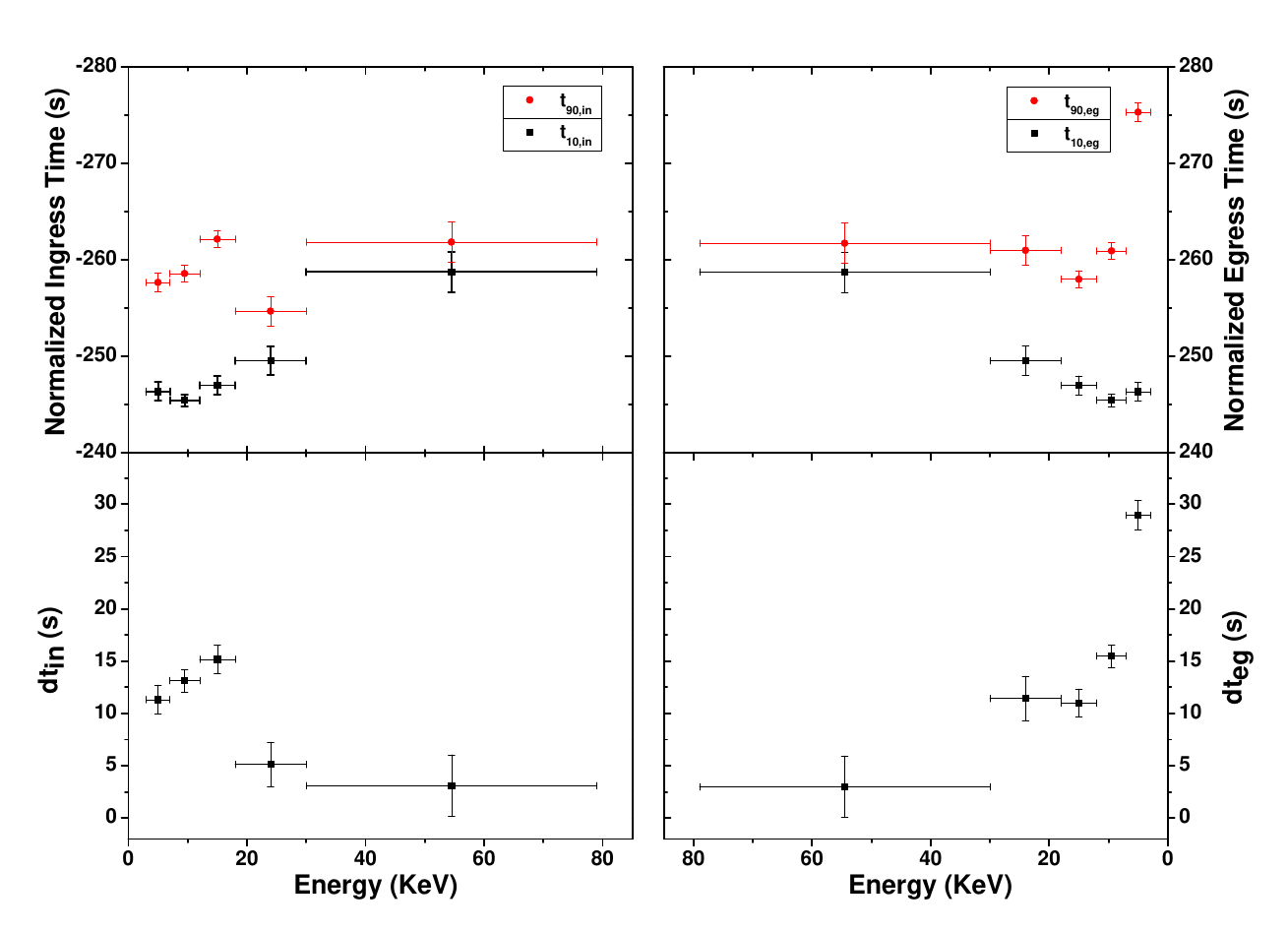}
    \caption{Measured eclipse times, $t_{90}$ and $t_{10}$, as functions of energy for the ingress (top left panel) and egress (top right panel) along with the variation of ingress duration $dt_{in}$ (bottom left panel) and egress duration $dt_{eg}$ (bottom right panel) with different energies.}
    \label{fig:10-90}
\end{figure*}

For a more detailed analysis, we further divided the ingress and egress phases into two-time frames, defined as \(t_{10}\) and \(t_{90}\). Here, \(t_{10}\) and \(t_{90}\) represent the points where the count rate reaches 10\% and 90\% of the average burst-free, out-of-eclipse count rate, respectively. For the ingress phase, \(t_{90}\) marks the onset of the count rate’s gradual decline, while \(t_{10}\) denotes its approach to the out-of-eclipse count rate level.\\

To account for random fluctuations, these times are defined as the points at which the average count rate first surpasses the specified percentage and remains there for at least five seconds \cite{knight2022eclipse}. The measured values of \(t_{10}\) and \(t_{90}\) across different energy ranges, from 3–7 keV to 30–79 keV, are displayed in the upper panels of Fig. \ref{fig:10-90}, showing ingress on the left and egress on the right, with the corresponding ingress/egress durations shown in the lower panels.\\

As seen in Fig. \ref{fig:10-90}, the ingress start time \(t_{90}\) increases for higher energy ranges from 3–7 keV to 12–18 keV, decreases for the 18–30 keV range, and then rises again in the 30–79 keV range. Similarly, the ingress end time \(t_{10}\) shows a slight decrease from 3–7 keV to 7–12 keV, followed by a linear increase with higher energy ranges. This pattern indicates that ingress begins later at higher photon energies but ends quickly, whereas the lower energy ranges show longer ingress durations, particularly within the 12–18 keV range, with durations decreasing towards the 3–7 keV region.\\

The observed behavior of the ingress start time $t_{90}$ and end time $t_{10}$ across different energy ranges in EXO 0748-676 reflects how the neutron star's environment and radiation characteristics vary with photon energy. As seen in Fig. \ref{fig:10-90}, ingress Start Time \(t_{90}\) decreases for higher energy ranges from 3–7 keV to 12–18 keV, increases for the 18–30 keV range, and then again decreases for the 30–79 keV range. The decrease in  \(t_{90}\) at higher photon energies, from 3–7 keV up to 12–18 keV, suggests that the high-energy X-rays encounter the obscuring material later in the ingress phase. Higher energy photons can penetrate the neutron star’s surrounding environment more effectively, making them visible longer as the eclipse starts. For the 18–30 keV range, the increase in  \(t_{90}\) may indicate a transitional energy regime where the photons interact more strongly with the denser regions near the neutron star, becoming obscured sooner. Similarly, the subsequent decrease in  \(t_{90}\) for the 30–79 keV range might indicate that high-energy X-rays, although affected by scattering or absorption processes, still penetrate more effectively than intermediate-energy X-rays. These high-energy photons are likely interacting with less dense material, allowing them to be visible longer at the start of the ingress. \\

Similarly, the ingress end time \(t_{10}\) shows a slight increase from 3–7 keV to 7–12 keV, followed by a linear decrease with higher energy ranges. The initial increase in \(t_{10}\) from 3–7 keV to 7–12 keV suggests that lower-energy photons are more susceptible to absorption by material in the eclipse, causing the count rate to drop faster and earlier in the process. And the linear decrease in \(t_{10}\) with increasing photon energy indicates that high-energy photons end the ingress phase more quickly. This is because high-energy photons can traverse the surrounding material more effectively, which results in a sharper and more rapid end to the ingress phase for higher energy bands. \\

The pattern suggests that the ingress duration varies with energy. At lower energies, X-rays are more likely absorbed or scattered by the material around the neutron star, resulting in a longer ingress period at the energy range of 3-7 KeV to 12-18 KeV. At higher energies, 18 to 79 KeV, photons are less affected by scattering and absorption, which shortens the ingress duration. The shorter ingress duration at high energies implies that these photons can more directly reflect the neutron star’s emissions and are obscured only by the densest parts of the surrounding material.\\

The egress phase shows an interesting variation across energy ranges, reflecting changes in the timing of when photons of different energies begin and end their exit from the eclipse. The egress start time (\(t_{10}\)) slightly decreases from 3–7 keV to 7–12 keV, indicating that the eclipse ends earlier for these slightly higher energies. This trend then reverses as \(t_{10}\) begins to increase linearly with energy, suggesting that for higher photon energies, the egress starts progressively later. This delay at higher energies could imply that these photons are initially blocked by denser material and only gradually become visible as the line of sight clears. Meanwhile, the egress end time (\(t_{90}\)) generally decreases from 3–7 keV to 12–18 keV, meaning the eclipse duration shortens with energy in this range, causing higher energy photons to be visible sooner. However, \(t_{90}\) then shows a slight increase in the 18–30 keV range and becomes relatively constant at higher energies (30–79 keV), possibly due to higher energy photons passing through remaining obstructions more effectively. \\

Examining the overall duration of egress, it tends to decrease with increasing energy, suggesting that higher-energy photons are less affected by residual obstructions as the eclipse ends. A slight increase in egress duration is observed in the 18–30 keV range, indicating more complex interactions between these mid-energy photons and the material surrounding the source. Overall, this pattern reveals that higher-energy photons experience a shorter egress duration, likely because they penetrate through the material more effectively as it clears from the line of sight. This energy-dependent behavior of the ingress/egress phase offers insights into the density distribution and composition of the material surrounding the LMXB, highlighting how various photon energies respond to different extents of absorption and scattering. \\

\begin{figure*}[h!]
    \centering
    \includegraphics[width=0.9\linewidth]{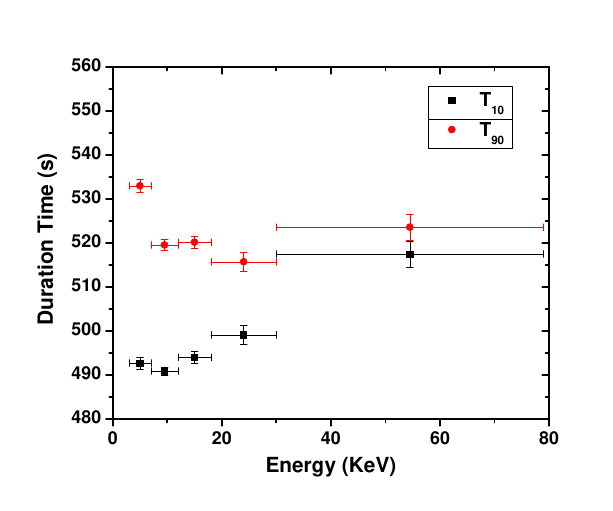}
    \caption{Variation of total eclipse duration $T_{10}$ and $T_{90}$ with different energy range for EXO 0748-676}
    \label{fig:ec_tot_dur}
\end{figure*} 

In the 3-79 keV band, the observed ingress and egress durations for EXO 0748-676 are 24.89 and 27.99 seconds, respectively. These durations reflect the time taken for the neutron star to enter and exit an eclipse, providing insight into the material surrounding the companion. Comparing this with a study conducted in the 0.2-10 keV range for the XMM-Newton observation of EXO 0748-676, it appears that eclipse behavior is energy-dependent, suggesting that an absorbing region surrounds the companion star and may be undergoing ablation. At lower energies, photons interact more with the absorbing material, while in higher-energy bands, such as 3-79 keV, the duration is shorter due to better photon penetration. This previous study also inferred a high neutron star mass and hard equation of state (EoS), suggesting the system’s compactness and gravitational effects influence these eclipse timings. Such comparisons across energy ranges allow for a more detailed understanding of the system's density, inclination, and mass properties \cite{knight2022eclipse}.\\

In our analysis of the total eclipse duration, characterized by \(T_{10}\) and \(T_{90}\), we observe distinct trends across various energy ranges, offering valuable insights into the dynamics of the eclipse. The total duration \(T_x\) is calculated as \(T_x = t_{x,\text{eg}} - t_{x,\text{in}}\), where \(x\) represents either 10 or 90. Figure \ref{fig:ec_tot_dur} illustrates the variation in total eclipse duration across different energy ranges. From Fig. \ref{fig:ec_tot_dur}, for \(T_{10}\), we observe a decrease from the energy range of 3-7 keV to 7-12 keV, suggesting that at these lower energy levels, the eclipse duration is shorter, indicating a rapid transition as the neutron star obscures the X-ray source. Beyond 12 keV, \(T_{10}\) gradually and linearly increases with energy. This could be due to increased complexity in interactions at higher photon energies, potentially related to variations in the density or structure of the material surrounding the neutron star.\\

In contrast, \(T_{90}\) exhibits a decrease from the 3-7 keV range to the 18-30 keV range, indicating that the onset of the gradual decline in count rate occurs earlier in these energy bands. This rapid decline in the observable count rate could be attributed to higher photon absorption or scattering processes occurring in the material surrounding the neutron star. After the 18-30 keV range, \(T_{90}\) begins to increase, suggesting that the eclipse duration becomes prolonged at higher energies. This could be due to increased photon penetration through the absorbing material or the effects of relativistic beaming, where high-energy photons are emitted in a more focused manner.\\

Furthermore, the difference \(T_{90} - T_{10}\) gradually decreases with increasing energy, indicating that as photon energy rises, the duration difference between the initial onset of the count rate decline and its approach to the out-of-eclipse level becomes smaller. This observation may suggest that higher-energy photons experience a more uniform interaction with the surrounding material, leading to a more consistent eclipse duration. These trends in the duration of the eclipse phases, influenced by energy levels, provide valuable information about the structure and behavior of the material surrounding the neutron star, including its density and composition.\\



\subsection{Phase-averaged Spectral analysis}
\label{spectral}


The X-ray spectral analysis of the source observed by NuSTAR's FPMA and FPMB instruments across the 3-79 keV energy range provides critical insights into the emission processes of this system. The analysis was performed using the \texttt{xspec} tool (v12.14.1) \cite{arnaud1996xspec} to model the observed spectrum, which was initially grouped to ensure a minimum of 20 counts per spectral bin using the \texttt{ftool GRPPHA}. This binning process improves the statistical significance of the spectral data by minimizing the impact of Poisson noise, particularly in regions with low photon counts. To account for the relative calibration differences between FPMA and FPMB, a \texttt{constant model} was applied, fixing FPMA to unity and allowing the FPMB normalization to vary.\\

\begin{figure*}[h!]
    \centering
    {\includegraphics[width=.9\linewidth]{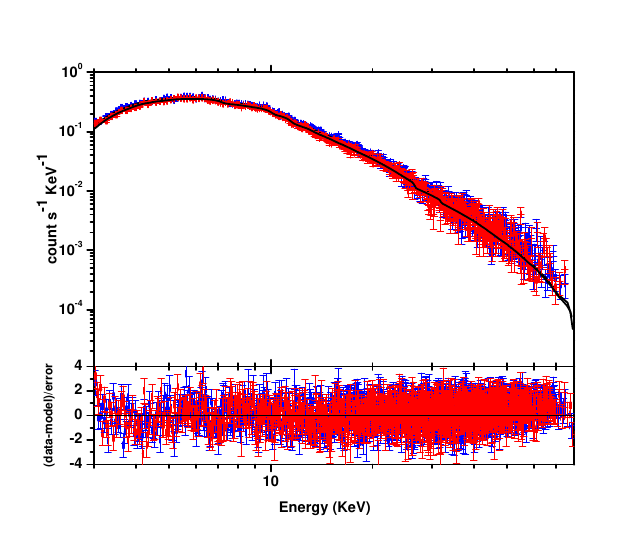}}
    \caption{The diagram represents an energy spectrum of NuSTAR observations in the 3-79 keV energy range. The bottom panel represents residuals of continuum model fit of \texttt{constant * phabs * cutoffpl} configuration. Red and Blue color corresponds to NuSTAR FPMA and FPMB spectra respectively.}
    \label{fig:spectrum}
\end{figure*}

\begin{table}[h!]
\centering
\renewcommand{\arraystretch}{1.5} 
\setlength{\tabcolsep}{12pt} 
\caption{Best fitted Spectral Parameters for EXO 0748-676 with 90\% Confidence Errors  and $\chi^2_{reduced}=1.006$ for model constant*phabs*cutoffpl}
\label{tab2}
\begin{tabular}{|c|c|c|c|c|}
\hline
\rowcolor[HTML]{D9EAD3} 
\textbf{Parameter} & \textbf{Component} & \textbf{Unit} & \textbf{Value} & \textbf{Confidence Range (90\%)} \\ \hline
$C$                & Constant           & -             & $1.00000$ (frozen)                       & -                                  \\ \hline
\rowcolor[HTML]{F4CCCC} 
$N_H$              & Phabs              & $10^{22}$     & $6.35318_{-0.24847}^{+0.24962}$          & $6.1047 \; - \; 6.60279$           \\ \hline
$\Gamma$           & CutoffPL           & -             & $1.23911_{-0.02228}^{+0.02218}$          & $1.21683 \; - \; 1.26129$          \\ \hline
\rowcolor[HTML]{FFF2CC} 
$E_{\text{cut}}$   & CutoffPL           & keV           & $36.2009_{-1.63141}^{+1.77375}$          & $34.5695 \; - \; 37.9746$          \\ \hline
$N$                & CutoffPL           &  $10^{-2}$           & $1.77416_{-0.0711}^{+0.0739}$ & $0.0170304 \; - \; 0.0184807$ \\ \hline
\end{tabular}
\end{table}

This study explores EXO 0748-676’s emission characteristics through two distinct spectral continuum models applied to data obtained from the NuSTAR telescope: the first continuum model (Model 1), which uses a \texttt{constant * phabs * cutoffpl} configuration, and the second continuum model (Model 2), a more complex model that incorporates \texttt{constant * phabs * highecut (powerlaw + bbodyrad)}. Each model reveals key physical insights into the binary system's structure and provides a comparative perspective on the mechanisms that govern its emission.\\

Model 1 uses a cutoff power-law to describe the spectrum, along with interstellar absorption modeled by \texttt{phabs}. This model represents the X-ray emission as a power-law with an exponential cutoff, suitable for characterizing Comptonized emission from an optically thin, high-temperature electron plasma located near the neutron star. The spectral fit resulted in reduced chi-squared values ($\chi^2_{\nu}$) of 0.982 for FPMA and 1.006 for FPMB, indicating that the model accurately describes the data, as illustrated in Fig. \ref{fig:spectrum}. Best fitted Spectral Parameters for EXO 0748-676 with 90\% Confidence Errors are listed in Table \ref{tab2}. The fitted value for the hydrogen column density ($N_H$) is \(6.35318_{-0.24847}^{+0.24962} \times 10^{22} \, \text{cm}^{-2}\), consistent with previous observations and indicating substantial absorption, likely due to both interstellar matter and circumstellar material around the binary. The power-law photon index ($\Gamma$) of \(1.23911_{-0.02228}^{+0.02218}\) suggests a hard spectrum, indicative of high-energy emissions from the electron plasma. Additionally, the high-energy cutoff parameter (\(E_{\text{cut}}\)) is found to be \(36.2009_{-1.63141}^{+1.77375} \, \text{keV}\), suggesting that most of the energy is emitted below this threshold, which points to the temperature of the corona surrounding the neutron star. Finally, the power-law normalization (\(N\)) of \(1.77416 _{-0.0711}^{+0.0739} \times 10^{-2}\) defines the relative flux of this component. Together, these parameters provide a reliable depiction of the system’s X-ray continuum emission, primarily shaped by the high-energy scattering processes occurring in the hot corona.\\

\begin{figure*}[h!]
    \centering
    {\includegraphics[width=.9\linewidth]{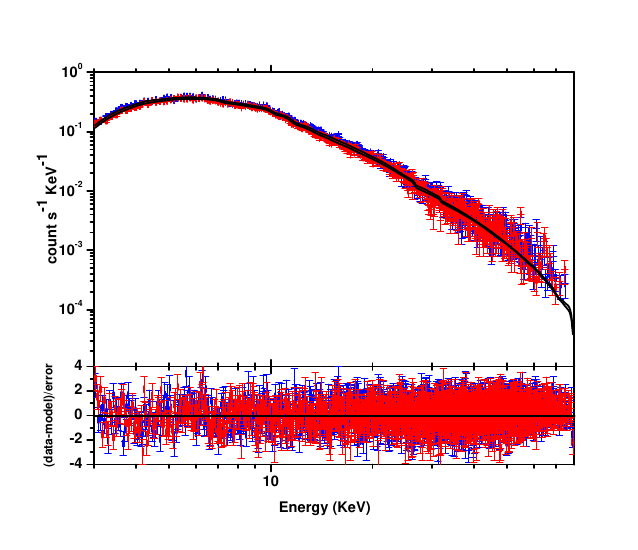}}
    \caption{The diagram represents an energy spectrum of NuSTAR observations in the 3-79 keV energy range. The bottom panel represents residuals of continuum model fit of \texttt{constant * phabs * cutoffpl} configuration. Red and Blue color corresponds to NuSTAR FPMA and FPMB spectra respectively.}
    \label{fig:spectrum2}
\end{figure*}

\begin{table}[h!]
\centering
\renewcommand{\arraystretch}{1.5} 
\setlength{\tabcolsep}{12pt} 
\caption{Fitted Spectral Parameters for EXO 0748-676 with 90\% Confidence Errors and $\chi^2_{reduced}=0.999$ for Model constant*phabs*highecut(powerlaw + bbodyrad)}
\label{tab3}
\begin{tabular}{|c|c|c|c|c|}
\hline
\rowcolor[HTML]{D9EAD3} 
\textbf{Parameter} & \textbf{Component} & \textbf{Unit} & \textbf{Value} & \textbf{Confidence Range (90\%)} \\ \hline
$C$                & Constant           & -             & $1.00000$ (frozen)                       & -                                  \\ \hline
\rowcolor[HTML]{F4CCCC} 
$N_H$              & Phabs              & $10^{22}$     & $4.39860_{-0.49738}^{+0.51237}$          & $3.90122 \; - \; 4.91097$          \\ \hline
$E_{\text{cutoff}}$& HighEcut           & keV           & $10.8491_{-0.58223}^{+0.64618}$          & $10.2669 \; - \; 11.4953$          \\ \hline
\rowcolor[HTML]{FFF2CC} 
$E_{\text{fold}}$  & HighEcut           & keV           & $28.4367_{-2.5571}^{+3.0009}$            & $25.8796 \; - \; 31.4376$          \\ \hline
$\Gamma$           & Powerlaw           & -             & $0.996201_{-0.09161}^{+0.08842}$         & $0.904596 \; - \; 1.08462$         \\ \hline
\rowcolor[HTML]{D9EAD3} 
$N_{\text{pl}}$    & Powerlaw           & $ 10^{-3}  $           & $6.76645_{-1.48}^{+1.82}$ & $0.00528726 \; - \; 0.00858597$ \\ \hline
$kT$               & BBodyRad           & keV           & $1.48324_{-0.06464}^{+0.06034}$          & $1.4186 \; - \; 1.54358$          \\ \hline
\rowcolor[HTML]{F4CCCC} 
$N_{\text{bb}}$    & BBodyRad           & -             & $0.645857_{-0.10863}^{+0.10647}$         & $0.537224 \; - \; 0.752324$       \\ \hline
\end{tabular}
\end{table}

Model 2 offers a more sophisticated interpretation, incorporating both non-thermal and thermal emission components, which gives a comprehensive view of the system. This model includes a combination of a power-law, similar to that in Model 1, with an additional blackbody radiation component representing thermal emission from the neutron star's surface or the inner accretion disk. Here, the cutoff and folding energies are modeled using the \texttt{highecut} component, which captures the high-energy break more flexibly. The spectral fitting yielded reduced chi-squared values of \( \chi^2_{\nu} = 0.979 \) for FPMA and \( \chi^2_{\nu} = 0.999 \) for FPMB, suggesting that the model effectively represents the data, as shown in Fig. \ref{fig:spectrum2}. The optimal spectral parameters for EXO 0748-676, along with their 90\% confidence errors, are detailed in Table \ref{tab3}.\\

The hydrogen column density (\(N_H\)) in this model is \(4.39860_{-0.49738}^{+0.51237} \times 10^{22} \, \text{cm}^{-2}\), which is slightly lower than in Model 1. The model’s cutoff energy (\(E_{\text{cutoff}}\)) is \(10.8491_{-0.58223}^{+0.64618} \, \text{keV}\), and the folding energy (\(E_{\text{fold}}\)) is \(28.4367_{-2.5571}^{+3.0009} \, \text{keV}\), parameters that characterize the Comptonized corona emission and indicate possible variations in the temperature and density of the electron cloud surrounding the neutron star. The power-law photon index (\(\Gamma\)) of \(0.996201_{-0.09161}^{+0.08842}\) suggests a particularly hard spectrum, indicative of strong Compton scattering effects within a highly energetic environment.\\

A notable feature of Model 2 is the inclusion of a blackbody radiation component. This component, with a temperature (\(kT\)) of \(1.48324_{-0.06464}^{+0.06034} \, \text{keV}\), reflects thermal emission, possibly originating from the neutron star’s heated surface due to accretion activity. This thermal component represents a blackbody surface temperature on the order of \(1.7 \times 10^7 \, \text{K}\), which provides insight into the neutron star's thermal processes and how accretion impacts the surface. The power-law normalization (\(N_{\text{pl}}\)) is \(6.76645 _{-1.48}^{+1.82}  \times 10^{-3}\), and the blackbody normalization (\(N_{\text{bb}}\)) is \(0.645857_{-0.10863}^{+0.10647}\), showing the respective contributions of each emission type within the spectrum. This model, with its thermal component, gives a more nuanced view of the spectral properties, separating out the potential emission sources, including the neutron star surface, the boundary layer, and the surrounding accretion disk.\\

The two models provide complementary insights into the behavior of EXO 0748-676. Model 1 employs a single-component approach that effectively captures the continuum emission, suggesting that the system's high-energy behavior is predominantly influenced by scattering processes within a hot, optically thin corona. In contrast, Model 2 offers a more nuanced exploration of the spectral composition by isolating both thermal and non-thermal contributions, indicating the likely presence of a corona alongside a heated neutron star surface. This two-component model points to a multi-temperature structure within the binary system, with the lower cutoff energy in Model 2 potentially reflecting cooler regions within the accretion environment or variations in accretion dynamics over time. Overall, both models yield valuable insights, with Model 1 providing a robust representation of continuum emission, while Model 2 enhances our understanding of the thermal and non-thermal processes involved. The ability of Model 2 to distinguish between these components is particularly relevant for elucidating the physical conditions surrounding the neutron star, including the interplay between accretion disk dynamics, coronal heating, and surface emission. Such spectral analysis is crucial for probing the physical conditions governing EXO 0748-676, illuminating its high-energy emissions, and contributing to a broader understanding of low-mass X-ray binary systems and neutron star characteristics. Further observations across different energy bands and over extended periods would be beneficial for refining these models and enhancing our comprehension of the spectral complexity and temporal variability inherent to EXO 0748-676.\\

We have also calculated the flux in a full energy range of 3-79 KeV using both the continuum models with 90\% confidence errors. The first model provided a flux of \(381.17^{+2.97}_{-2.97} \times 10^{-12}\, \text{ergs/cm}^2/\text{s}\). This model suggests a significant contribution from non-thermal processes, indicating the influence of high-energy emissions in the observed spectrum. Conversely, the second model yielded a slightly lower flux of \(379.56^{+2.04}_{-1.90}\times 10^{-12}\, \text{ergs/cm}^2/\text{s}\). The minor difference in flux between the two models highlights the variability in the spectral shape and the contributions of thermal and non-thermal emissions. Overall, both models underscore the complexity of the accretion processes and emission mechanisms present in EXO 0748-676, providing crucial insights into the neutron star's environment and the underlying physics governing low-mass X-ray binaries.\\

The spectral fits obtained from NuSTAR observations were compared with earlier results from the \texttt{IBIS/ISGRI} instrument on the INTEGRAL satellite, which observed EXO 0748-676 in the 20-100 keV range in 2003 and 2004 \cite{gotz2006integral}. These earlier observations showed slightly softer photon indices compared to NuSTAR’s, with values of $ \Gamma = 1.3 \pm 0.4$ in 2003 and $ \Gamma = 1.6 \pm 0.4$ in 2004. This indicates that the emission was somewhat softer during those periods. Softer photon indices typically suggest that the spectrum contains more lower-energy photons, implying differences in the energy distribution of the emitted radiation between the observations. In contrast, NuSTAR’s observations revealed a harder spectrum, suggesting a more compact and energetic emission region.\\

In addition to the differences in photon indices, the cutoff energies observed by INTEGRAL in 2003 and 2004 were 44 keV and 50 keV, respectively. These values are higher than NuSTAR’s observed cutoff energy of 36.2 keV, which suggests that the Comptonizing region or corona was cooler during the NuSTAR observations. A lower cutoff energy means that the high-energy photons are less frequent, which could point to a less energetic particle population responsible for upscattering the photons. The differences in cutoff energies could be linked to variations in the accretion environment or changes in the physical conditions around the neutron star at different times.\\

The flux values observed by IBIS/ISGRI varied over time, with measurements of \(320 \times 10^{-12} \, \text{erg/cm}^2/\text{s}\) in 2003 and \(421 \times 10^{-12} \, \text{erg/cm}^2/\text{s}\) in 2004. These values reflect the typical fluctuations seen in accretion-driven systems, where changes in the accretion rate can lead to variations in the observed flux. In comparison to the 2004 data, the flux values obtained from NuSTAR’s spectral fits are lower. For Model 1, the flux was \(381.17^{+2.97}_{-2.97} \times 10^{-12} \, \text{ergs/cm}^2/\text{s}\), while for Model 2, the flux was slightly lower at \(379.56^{+2.04}_{-1.90} \times 10^{-12} \, \text{ergs/cm}^2/\text{s}\). These flux values indicate that EXO 0748-676 was dimmer during the NuSTAR observations compared to the earlier measurements from 2004. The decrease in flux may suggest reduced accretion activity or changes in the system's viewing angle.\\

To further understand the physical properties of EXO 0748-676, the distances to the source were considered. Several distance estimates have been proposed in the literature. Özel et al. suggested a distance of 9.2 $\pm$ 1.0 kpc based on the gravitational redshift of the system \cite{ozel2006soft}. Wolff et al. derived a distance of 7.7 kpc from observations of a helium burst \cite{wolff2005strong}, while Galloway et al. analyzed several type I X-ray bursts and estimated a distance of 7.4 kpc \cite{galloway2008thermonuclear}. More recently, Galloway, Özel, and Psaltis proposed a distance of 7.1 $\pm$ 1.2 kpc based on the touchdown flux and the high binary inclination of the system \cite{galloway2008biases, zhang2011distance}. These varying distance estimates impact the calculation of the source’s luminosity, which depends on both the flux and the distance.

\begin{table}[h!]
\centering
\renewcommand{\arraystretch}{1.5} 
\setlength{\tabcolsep}{8pt} 
\caption{Luminosity for EXO 0748-676 Models with Flux Values: Model 1 Flux = \(381.17^{+2.97}_{-2.97} \times 10^{-12}\) ergs/cm$^2$/s, Model 2 Flux = \(379.56^{+2.04}_{-1.90}\times 10^{-12}\) ergs/cm$^2$/s}
\label{tab4}
\begin{tabular}{|c|c|c|c|c|}

\hline
\rowcolor[HTML]{D9EAD3} 
\textbf{Distance (kpc)} & \textbf{Model 1 Luminosity (ergs/s)} & \textbf{Model 2 Luminosity (ergs/s)} \\ \hline

9.2 ± 1 & \(3.86^{+0.92}_{-0.82} \times 10^{36}\) & \(3.84^{+0.91}_{-0.85} \times 10^{36}\) \\ \hline

\rowcolor[HTML]{F4CCCC} 
7.7 ± 0.9 & \(2.70^{+0.70}_{-0.61} \times 10^{36}\) & \(2.69^{+0.69}_{-0.70} \times 10^{36}\) \\ \hline

7.4 & \(2.50^{+0.02}_{-0.02} \times 10^{36}\) & \(2.49^{+0.13}_{-0.01} \times 10^{36}\) \\ \hline

\rowcolor[HTML]{FFF2CC} 
7.1 ± 1.2 & \(2.30^{+0.87}_{-0.13} \times 10^{36}\) & \(2.29^{+0.79}_{-0.14} \times 10^{36}\) \\ \hline

\end{tabular}
\end{table}

Using the flux values obtained from both models, the luminosity of EXO 0748-676 was calculated for each of the proposed distances of 9.2 kpc, 7.7 kpc, 7.4 kpc, and 7.1 kpc. The calculated luminosity are detailed in Table \ref{tab4}. These results demonstrate a clear relationship between the distance and the luminosity of EXO 0748-676, with larger distances corresponding to higher luminosities. This relationship is a direct consequence of the inverse square law, where luminosity increases with the square of the distance for a given flux. The difference between the two models is relatively small, with Model 1 consistently yielding slightly higher luminosity values than Model 2. However, the uncertainty in the flux from Model 2 is larger, leading to broader error margins in the derived luminosities.

When comparing these results to previous luminosity estimates, we find consistency in the overall trends, though there are some notable differences. Díaz Trigo et al. \cite{trigo2006spectral} reported an absorbed flux of \(2.25 \times 10^{-10} \, \text{erg/cm}^2/\text{s}\) for EXO 0748-676 in the 0.6–10 keV range, when corrected for no attenuation, which corresponds to an unabsorbed flux of \(2.81 \times 10^{-10} \, \text{erg/cm}^2/\text{s}\). This flux value is lower than the fluxes obtained from both NuSTAR models. Boirin et al. \cite{boirin2007discovery} further calculated an unabsorbed bolometric flux of \(8.44 \times 10^{-10} \, \text{erg/cm}^2/\text{s}\) across the 0.1-100 keV energy band which is greater than that predicted at NuSTAR. However, when considering a distance of 5 kpc, their data gives a luminosity of \(2.52 \times 10^{36} \, \text{erg/s}\), which is comparable to the luminosities derived from both NuSTAR models at 7.4 kpc.

When comparing these results with Díaz Trigo's estimates of \(8.44 \times 10^{35} \, \text{erg/s}\) at 5 kpc and \(3.36 \times 10^{36} \, \text{erg/s}\) at 10 kpc in the 0.6-10 keV range, we observe that our findings are in line with these estimates, particularly when considering the range of distances. Specifically, our luminosity results suggest that the pulsar is relatively bright at the higher distance of 9.2 kpc, which aligns with the higher end of the luminosities reported by Díaz Trigo.\\

The luminosities derived from both NuSTAR models are consistent with previous studies when taking into account the different distance estimates and the variability in the accretion environment of EXO 0748-676. The slightly higher luminosity values obtained from Model 1 reflect the marginally higher fluxes observed with this model, while Model 2 introduces greater uncertainty in the results due to its broader error margins. Overall, the results provide valuable insight into the emission properties of EXO 0748-676 and reinforce the importance of accurate distance measurements in determining the physical characteristics of such systems, as variations in distance can significantly affect luminosity calculations. The discrepancies between our estimates and those reported by Boirin et al. at different distances may stem from differences in energy ranges analyzed or variations in the flux values used in their calculations. Furthermore, our findings highlight the importance of using multiple distance estimates to obtain a comprehensive understanding of the source's intrinsic properties, as the luminosity values can provide insights into the physical processes occurring in EXO 0748-676, particularly during burst events. Overall, the luminosity derived from our analysis demonstrates that EXO 0748-676 remains a key target for understanding the behavior of low-mass X-ray binaries, contributing to the ongoing discourse regarding the physical mechanisms at play during X-ray outbursts.\\


\section{Conclusion}
\label{conc}

The analysis of Type I X-ray bursts from the low-mass X-ray binary EXO 0748-676, based on detailed NuSTAR observations, provides critical insights into the mechanisms behind thermonuclear explosions on neutron stars. This study examined the light curve data, revealing the dynamic nature of the accretion process and the conditions necessary for hydrogen and helium ignition. A prominent Type I X-ray burst observed at \(X = 18,479.97\), with a peak rate of 1,615.88 counts per second, highlights the intense thermonuclear activity caused by the unstable burning of accreted material on the neutron star's surface.\\

The findings demonstrate that the thermonuclear ignition is most pronounced in mid-energy bands (7-18 keV), while the cooling phase is dominated by soft X-rays (3-7 keV), underscoring the energy-dependent nature of these bursts. The intervals between bursts, including a 984-second gap between the primary and secondary bursts, provide insights into the accretion dynamics and material instabilities. Additionally, the correlation between the timing of burst events and the neutron star's optical period suggests a close relationship between accretion from the companion star and burst modulation.\\

The analysis of the eclipse profile of EXO 0748-676 revealed that higher-energy photons penetrate surrounding material more effectively, as indicated by energy-dependent ingress and egress durations. Variations in \(T_{10}\) and \(T_{90}\) across energy bands reflect the density structure of the environment, revealing complex interactions between X-ray emissions and the material surrounding the neutron star. The study also applied two spectral continuum models to describe the system’s X-ray emissions. Model 1 focused on Comptonization in a high-temperature electron plasma, while Model 2 introduced a dual-component approach, emphasizing both thermal and non-thermal processes. The close agreement between the models strengthens the reliability of the analysis, though differences point to complexities in the emission mechanisms.\\

When compared to earlier observations, this study suggests temporal changes in the accretion environment of EXO 0748-676. The NuSTAR data revealed a harder spectrum and lower cutoff energies, indicating a cooler Comptonizing region during the observation period. These findings emphasize the dynamic nature of the system and its variable emission characteristics.\\

In the broader context of neutron star behavior, the results support the idea that neutron stars in low-mass X-ray binaries, such as EXO 0748-676, could exhibit characteristics similar to millisecond pulsars. These neutron stars often rotate rapidly and may emit periodic pulses of radiation due to interactions between their magnetic fields and accretion flows. Although the primary focus of this study was on X-ray bursts and spectral characteristics, the presence of strong magnetic fields and the potential for pulsar-like emissions are important factors in understanding the behavior of such systems.\\

Overall, this comprehensive study deepens the understanding of X-ray bursts, accretion dynamics, and spectral properties in neutron star systems. Future research, including more pulsar-focused studies, will be essential for further exploring the interactions between accretion, nuclear burning, and the neutron star’s magnetic environment. These investigations will contribute to the broader understanding of neutron star evolution, stellar processes, and high-energy astrophysical phenomena.\\

\section{Data Availability}
\label{availability}

The data utilized in this study is available to the public through the HEASARC data archive, allowing for research purposes.

\bibliographystyle{unsrtnat}
\bibliography{references}  






\end{document}